\documentclass[12pt,preprint]{aastex}

\newcommand{\be} {\begin{equation}}

\def\rrat {RRAT\,J1819--1458\,}

\newcommand{\CXO}{{\it Chandra}\,}

\newcommand{\XMM}{{\em XMM--Newton}}

\newcommand{\bc}{\begin{center}}
\newcommand{\ec}{\end{center}}
\def\ltsima{$\; \buildrel < \over \sim \;$}
\def\lsim{\lower.5ex\hbox{\ltsima}}
\def\loe{\lower.5ex\hbox{\ltsima}}
\def\gtsima{$\; \buildrel > \over \sim \;$}
\def\gsim{\lower.5ex\hbox{\gtsima}}
\def\goe{\lower.5ex\hbox{\gtsima}}

\def\ltsima{$\; \buildrel < \over \sim \;$}
\def\lsim{\lower.5ex\hbox{\ltsima}}
\def\loe{\lower.5ex\hbox{\ltsima}}
\def\gtsima{$\; \buildrel > \over \sim \;$}
\def\gsim{\lower.5ex\hbox{\gtsima}}
\def\goe{\lower.5ex\hbox{\gtsima}}

\def\ergs {erg\,s$^{-1}$}
\def\ergscm2 {erg\,s$^{-1}$cm$^{-2}$}

\def\cm2 {cm$^{-2}$}
\def\arcsec{$^{\prime\prime}$}

\shortauthors{REA ET AL.}
\shorttitle{Extended X-ray emission around \rrat}

\begin{document}
\title{Discovery of extended X-ray emission around the highly magnetic \rrat}

\author{N. Rea\altaffilmark{1}, M.~A. McLaughlin\altaffilmark{2,3,4}, B.~M. Gaensler\altaffilmark{5}, P.~O. Slane\altaffilmark{6}, L. Stella\altaffilmark{7}, \\ S.~P. Reynolds\altaffilmark{8}, M. Burgay\altaffilmark{9}, G.~L. Israel\altaffilmark{7}, A. Possenti\altaffilmark{9}, S. Chatterjee\altaffilmark{10}}
\altaffiltext{1}{Astronomical Institute "Anton Pannekoek",
University of Amsterdam, Science Park 904, Postbus 94249, 1090 GE, Amsterdam, The
Netherlands; n.rea@uva.nl}
\altaffiltext{2}{Department of Physics, West Virginia University, Morgantown, WV 26501, USA}
\altaffiltext{3}{National Radio Astronomy Observatory, Green Bank, WV 24944, USA}
\altaffiltext{4}{Alfred P. Sloan Research Fellow}
\altaffiltext{5}{Sydney Institute for Astronomy, School of Physics, The University of Sydney, NSW 2006, Australia}
\altaffiltext{6}{Harvard-Smithsonian Center for Astrophysics, 60 Garden Street, Cambridge, MA 02138, USA}
\altaffiltext{7}{INAF-Osservatorio Astronomico di Roma, via Frascati 33, I-00040 Monteporzio Catone, Italy}
\altaffiltext{8}{Department of Physics, North Carolina State University, P.O. Box 8202, Raleigh, NC 27695, USA}
\altaffiltext{9}{INAF-Osservatorio Astronomico di Cagliari, Loc. Poggio dei Pini, Strada 54, 09012 Capoterra, Italy}
\altaffiltext{10}{Department of Astronomy and National Astronomy and Ionosphere Center, Cornell University, Ithaca, NY 14853-6801, USA}

\begin{abstract}

  We report on the discovery of extended X-ray emission around the
  high magnetic field Rotating Radio Transient J1819--1458. Using a
  30\,ks \CXO\, ACIS-S observation, we found significant evidence for
  extended X-ray emission with a peculiar shape: a compact region out
  to $\sim$5.5\arcsec, and more diffuse emission extending out to
  $\sim$13\arcsec\, from the source. The most plausible interpretation is
  a nebula somehow powered by the pulsar, although the small number of
  counts prevents a conclusive answer on the nature of this
  emission. \rrat's spin-down energy loss rate ($\dot{E}_{\rm
    rot}\sim3\times10^{32}$\ergs) is much lower than that of other
  pulsars with observed spin-down powered pulsar wind nebulae (PWNe),
  and implies a rather high X-ray efficiency of $\eta_{X}\equiv L_{\rm
    pwn; 0.5-8\,keV}/\dot{E}_{\rm rot}\sim0.2$ at converting spin-down
  power into the PWN X-ray emission. This suggests the need of an
  additional source of energy rather than the spin-down power alone,
  such as the high magnetic energy of this source. Furthermore, this
  \CXO\, observation allowed us to refine the positional accuracy of
  \rrat\, to a radius of $\sim$0.3\arcsec, and confirms the presence of
  X-ray pulsations and the $\sim$1\,keV absorption line, previously
  observed in the X-ray emission of this source.

\end{abstract}


\section{INTRODUCTION}
\label{intro}

All pulsars lose energy through a wind of relativistic particles.
This relativistic wind has long been recognized to shock against the
ambient medium, giving rise to ``nebulae'' powered by the pulsars,
which usually emit at X-ray and radio wavelengths.  It can be expected
that all pulsars will be surrounded by such nebulae, usually powered by
the pulsar rotational energy. The brightness and shapes of these
nebulae depend on the pulsar properties, on the ambient medium density
and anisotropies, and on the pulsar proper motion. In the most
idealized case, an isotropic wind would form a spherical termination
shock (Kennel \& Coroniti 1984).  The superb angular resolution of
\CXO\, has resulted in the detection of roughly 50 nebulae, and has
shown that this idealized description is not sufficient to describe
most of their structures (Gaensler \& Slane 2006; Kargaltsev \& Pavlov
2008).  In general, these nebulae are anisotropic, show equatorial and
polar outflows, and have rich spectral structures. Most of the nebulae
detected thus far are associated with pulsars with high spin-down
energy loss rates ($\dot{E}_{\rm rot} > 10^{33}$\,ergs), the so called
spin-down powered pulsar wind nebulae (PWNe), but there are a few
cases where a nebula is observed but no pulsar is detected. The exact
physical origin and acceleration mechanism of the high-energy
particles in the pulsar winds are poorly understood, and not all
nebulae can be easily explained as spin-down powered PWNe.

Rotating radio transients (RRATs) are peculiar neutron stars which,
unlike normal radio pulsars, are detectable only through their
sporadic radio bursts (McLaughlin et al. 2006). We currently know of
over 20 of these objects (Deneva et al. 2008; Keane et al. 2009); they show a rather
broad range of spin-down properties, with inferred surface dipole
magnetic field strengths ranging from
$2\times10^{12}-5\times10^{13}$~G and characteristic ages ranging from
0.1--4~Myr. No sign of a companion star has yet been found for any of
these objects.

\rrat\, shows the most extreme and varied phenomenology of all the
RRATs, and is the best studied object in the class. Radio bursts are
detected every $\sim$3~minutes, and two glitches have been observed
which showed anomalous post-glitch recovery (Lyne et
al. 2009). \rrat\, has a 4.3\,s spin period, a characteristic age of
117~kyr and a spin-down energy loss rate $\dot{E}_{\rm rot}$ of
$3\times10^{32}$\,\ergs. The inferred surface dipolar magnetic field
of $B\sim5\times10^{13}$\,G is slightly higher than the electron
critical magnetic field of $B_{\rm cr}=4.4\times10^{13}$\,G,
suggesting a possible relationship with magnetars. By using the
dispersion measure and the Cordes \& Lazio (2002) electron density
model, the distance of \rrat\, is estimated to be
3.6~kpc. Furthermore, it is the only RRAT for which an X--ray
counterpart has been discovered (Reynolds et al.~2006; McLaughlin et
al.~2007; see also Rea \& McLaughlin 2008 for other non-detections).
After the serendipitous \CXO\, discovery of its X-ray emission, we
performed a 43\,ks \XMM\, observation in mid-2006 (McLaughlin et
al.~2007). This observation resulted in the detection of strong X--ray
pulsations at the radio period, with a $\sim$34\% pulsed fraction, and
a sinusoidal X-ray pulse profile aligned in phase with the radio
bursts. This observation also showed that the spectrum was well
modeled by an absorbed blackbody ($N_H = 3.8(2)\times 10^{21}$\,\cm2 ,
$kT =0.14(1)$\,keV, using Anders \& Grevesse (1989) solar abundances)
and a hint of a power-law component with $\Gamma\sim2$. Furthermore, a
broad spectral absorption line at $\sim$\,1\,keV was discovered, and
interpreted as either due to resonant cyclotron scattering, to the
neutron star atmosphere, or (less likely) to an overabundance of Ne
along the line of sight. The unabsorbed flux is
$1.5\times10^{-13}$\,\ergscm2 (0.3--5~keV). This converts to an X-ray
luminosity of $L_X \sim 4\times10^{33} (d/3.6~{\rm kpc})^2$\,\ergs ,
more than one order of magnitude higher than the spin-down luminosity.
Unfortunately, the \XMM\, observation lacked the angular resolution to
place meaningful constraints on small-scale extended emission.

We present here the results of a new \CXO\, observation of \rrat. The
observation and data reduction are reported in \S\ref{obs}, the
analysis and results in \S\ref{results}, and a discussion in
\S\ref{discussion}.

\section{OBSERVATION AND DATA REDUCTION}
\label{obs}

The \CXO\, X-ray Observatory observed \rrat\, for $\sim$30\,ks with
the Advanced CCD Imaging Spectrometer (ACIS) instrument on 2008 May 31
(ObsID\,7645) from 13:33:47 to 22:27:04 (UT) in {\tt VERY FAINT (VF)}
timed exposure imaging mode.  We used a 1/8 subarray, which provides a
time resolution of 0.4\,s, and the typical ACIS-S imaging and spectral
information. The source was positioned in the back-illuminated ACIS-S3
CCD at the nominal target position. Standard processing of the data
was performed by the \CXO\, X-ray Center to Level 1 and Level 2
(processing software DS 7.6.11.6). The data were reprocessed using the
CIAO software (version 4.0). We used the latest ACIS gain map, and
applied the time-dependent gain and charge transfer inefficiency
corrections. The data were then filtered for bad event grades and only
good time intervals were used. No high background events were
detected, resulting in a final exposure time of 27.88\,ks.

\section{ANALYSIS AND RESULTS}
\label{results}

\subsection{Accurate position}
\label{position}

We applied the {\tt wavedetect} tool to the ACIS-S cleaned image, and
found two X-ray bright stars in the field detected at a significance
of $>4\sigma$: \rrat\, at RA=18:19:34.173 and Dec=--14:58:03.57
(J2000), and another source at RA=18:19:32.36 and Dec=--14:57:58.67
(J2000), with statistical error circles of 0.01\arcsec\,and
0.18\arcsec\, radii, respectively (see also Fig.\,1; all uncertainties
in the text are reported at a 1$\sigma$ confidence level).  The latter
source is consistent with the
2MASS\footnote{http://www.ipac.caltech.edu/2mass/} star
18193233--1457584, which has a position of RA=18:19:32.34,
Dec=--14:57:58.5 (with an accuracy of 0.08\arcsec\, radius).

We could then perform a bore-site correction of the field to refine
the \rrat\, position and error circle. There were no problems with the
aspect solution during the observation. In particular, the
2MASS position of the source lies within the statistical 1$\sigma$
uncertainty of the serendipitous X-ray source, and that the next 2MASS
source is 8\arcsec\, away. Assuming that the association between the
2MASS star and the serendipitous X-ray source is sound, the final
\rrat\, position is RA=18:19:34.173 and Dec=--14:58:03.57, with a
1$\sigma$ associated error circle of 0.28\arcsec\, radius (derived
doing a quadratic mean of all statistical errors plus the 2MASS
catalogue intrinsic systematic error). This is the most accurate
position for \rrat, reported thus far, more accurate than that
achievable through radio timing (Lyne et al. 2009).

\subsection{Timing and spectroscopy}

For the timing and spectral analysis we extracted the source photons
from a circular region with 2.5\arcsec\, radius. Circular background
regions of radii 2.5\arcsec\, and 18\arcsec, far from the source, were
used for the timing and spectral analysis, respectively. \rrat\, has
an ACIS-S 0.3--10\,keV count rate of $0.041\pm0.001$ counts/s
(background subtracted).

For the timing analysis we corrected the arrival time of each photon
to the barycenter of the Solar System (using the {\tt JPL-DE405}
ephemeris). Using the {\tt Xrons} package, we folded the X-ray data
with the radio ephemeris (Lyne et al. 2009), revealing a sinusoidal
X-ray modulation with a 0.3--5\,keV pulsed fraction of 37$\pm$3\%,
defined as $(F_{\rm max}-F_{\rm min})/(F_{\rm max}+F_{\rm min})$, with
$F_{\rm max}$ and $F_{\rm min}$ the maximum and minimum counts of the
X-ray pulse profile.

The source spectrum was rebinned so as to have at least 25 counts per
spectral bin. We modeled the spectrum using the {\tt XSPEC} v.12.1
analysis package. We tried several single component continuum
models. The best fit was found with an absorbed blackbody plus an
absorption line which we modeled with a Gaussian function. Our best
fit values are: $N_{H}=(6\pm2)\times10^{21}$\cm2 and
$kT=0.12\pm0.02$\,keV for the continuum, and E$_G=1.0\pm0.2$\,keV,
$\sigma_G=0.12\pm0.06$\,keV, and an equivalent width EQW$=
103\pm25$\,eV for the absorption line ($\chi_{\nu}^2=1.02$ (29 dof);
see Fig.~2 left panel). The 0.3--5\~keV absorbed flux is $F_X =
(1.3\pm0.2)\times10^{-13}$\ergscm2 , while the inferred blackbody
radius is $2.1\pm0.4$\,km (assuming a 3.6\.kpc distance), smaller than
the whole neutron star surface, in accordance with the relatively high
pulsed fraction of this X-ray emission.

The pulse profile shape, pulsed fraction, spectral parameters, and
flux are all consistent, within the errors, with past measurements
(Reynolds et al. 2006; McLaughlin et al. 2007). Therefore, this new
\CXO\, observation did not provide any evidence for long term
variability. Likewise, no bursts nor aperiodic variations in the X-ray
flux were detected over the course of the observation.

Given the paucity of counts we cannot study the spectral line in more
detail, and we refer to McLaughlin et al (2007) and Rea et al. (2009
in preparation) for detailed studies based on \XMM\, data. Note that
the detection of the $\sim$1\,keV absorption line with \CXO\, shows
that the \XMM\, detection was not due to any instrumental effects and
confirms the reality of this feature.

\subsection{Imaging}
\label{imaging}

The very high-angular resolution of \CXO\ allowed us to perform for
the first time an image analysis on angular scales of a few arcseconds
(in the previous \CXO\, observation the source was off-axis, affording
a limited angular resolution; Reynolds et al. 2006). In Fig.~1 (left)
we show the image of the \rrat\, field in the 0.3--10\,keV energy
bands. Extended emission with a complex shape is apparent. There is a
more compact emission region extending to radii of roughly 5.5\arcsec,
from the pulsar, and then broader diffuse emission extending out to
13\arcsec. In Fig.~1 (right) we show the VLT-NACO K$_s$ field of
\rrat\, (Rea et al. 2009), which shows that this extended emission
cannot be due to the X-ray emission of a cluster of massive stars in
the line of sight. To infer the significance and estimate the
luminosity of the whole extended emission we built a {\it Chart}/{\it
  MARX} Point Spread Function (PSF) using the \rrat\, spectrum, and
subtracted it from the total 0.3--10\,keV cleaned image. From the
resulting image we extracted all the photons within a 13\arcsec\,
radius (this roughly corresponds to an extraction from an annular
region of 2.5--13\arcsec radii: see Fig.~1 left and Fig.\,2 right),
and we subtracted from it the background extracted from a similar
region far from the source (but in the same S3 CCD). We ended up with
an excess of 120$\pm$17 counts, which corresponds to a detection
significance of $\simeq 7\sigma$. The mean count rate for the whole
extended emission within 13\arcsec\, around the source is then
$(4.3\pm0.6)\times10^{-3}$ counts/s in the 0.3--10\,keV energy
band. Of these counts, 41$\pm$9 come from the 5.5\arcsec compact
region, with a mean count rate of $(1.5\pm0.3)\times10^{-3}$ counts/s.

In Fig.\,2 (right) we compare the background-subtracted surface
brightness radial distribution of our \CXO\, observation of \rrat\,
with that of the {\em Chart/MARX} PSF plus a simulated background
image. Both surface brightnesses were obtained by extracting counts from
50 annular regions (2 pixels wide each) centered on the source
position, and for the \rrat\, one, after removal of the serendipitous
point source and subtracting the background. This figure shows that
extended emission becomes detectable around 5 pixels
($\sim$2.5\arcsec) from the peak of the source PSF.  We also performed
the same analysis on a Level 2 event file which was re-built turning
off the pixel randomization, and applying the background cleaning of
the {\tt VF} mode. This neither changed the results nor improved their
significance.

Due to the small number of counts, the spectrum of the diffuse
emission is very poorly determined. With 120 counts in the
0.3--10\,keV range, we attempted a spectral modelling with a
power-law, which gave a good $\chi_{\nu}^2=1.07$ (6 dof; see Fig.\,2
left; the spectrum was grouped with 15 counts per bin). However, due
to the low number of counts the spectral parameters are poorly
constrained ($N_{H}<7\times10^{21}$\cm2 , $\Gamma=3.0\pm1.5$, and a
0.3--5\,keV absorbed flux of $(1.6\pm0.5)\times10^{-14}$\ergscm2 ). A
blackbody fit was also acceptable ($N_{H}<5\times10^{21}$\cm2 ,
$kT=0.21\pm0.12$\,keV), giving a $\chi_{\nu}^2=1.04$ (6 dof). However,
the blackbody fit showed systematic departures from the data at high
energies, and we therefore favor the power-law spectral model. We
performed the same fits keeping the $N_{H}$ fixed at the \rrat\,
value, finding consistent spectral parameters.

\section{DISCUSSION}
\label{discussion}

To date pulsars with observed X-ray nebulae have rotational power
$\dot{E}_{\rm rot}$ ranging from $10^{33-39}$\ergs (see Fig.\,3), and
they are usually rather young ($\tau_c \sim0.6-30$\,kyr). In this
respect, interpreting the nebula we see around \rrat\, as a spin-down
powered PWN is difficult given its low rotational power ($\dot{E}_{\rm
  rot}\simeq3\times10^{32}$\ergs) and its age ($\tau_c =
117$\,kyr). From the flux of the $\sim$13\arcsec\, extended emission
(see \S\ref{imaging}), and assuming a 3.6\,kpc distance, correcting
for absorption and extrapolating the flux in the 0.5--8\,keV energy
range, we then infer an X-ray efficiency $\eta_{X} \equiv L_{\rm pwn;
  0.5-8keV}/\dot{E}_{\rm rot} \sim 6\times10^{31}/3\times10^{32}
\simeq 0.2$ for transferring spin-down power to the X-ray PWN. This
X-ray efficiency is relatively high compared to the typical
$\eta_{X}\sim10^{-6}-10^{-1}$ observed in other pulsars showing PWNe
(see Fig.\,3; and also Cheng, Taam \& Wang 2004; Gaensler \& Slane
2006; Kargaltsev \& Pavlov 2008).

One possibility to explain the high X-ray efficiency of this putative
PWN might be that the distance is much closer than that inferred from
the radio DM of \rrat. If the real distance is e.g. half of the
current value (i.e. 1.8\,kpc), the X-ray efficiency would be
$\sim0.05$, similar to many other pulsars. However, even assuming a
smaller distance, the luminosity of this PWN exceeds the upper bound
trend for normal pulsars, $log L_{\rm pwn; 0.5-8keV}=1.6 log
\dot{E}_{\rm rot} - 24.2$ (Kargaltsev \& Pavlov 2008), which would
instead predict a luminosity of $L_{\rm pwn;
  0.5-8keV}\sim5.6\times10^{27}$\ergs, very difficult to reconcile
with our result, even allowing for a wrong distance for \rrat.

A second possibility might be that the compact 5.5\arcsec\, structure
we observe is a bow-shock nebula due to the pulsar moving
supersonically through the ambient medium. These types of nebulae are
indeed more commonly observed in older pulsars. In this scenario, the
larger scale extended emission could be part of the remnant of the
supernova explosion which formed \rrat. If we interpret the
5.5\arcsec\, structure as the termination radius of a bow-shock, we
get $R_{\rm TS} = 2.6\times10^{17}$\,cm (similar to other pulsars;
Kargaltsev \& Pavlov 2008), from which we infer a projected velocity
for \rrat\, of $v_{\rm p}\sim3\times10^{16}\dot{E}_{\rm
  rot}^{1/2}n^{-1/2}R_{\rm TS}^{-1}=20$\,km/s, rather small for a
pulsar showing a bow-shock considering a reasonable ambient medium
density (we assumed here {\it n}=1\,cm$^{-3}$ and a 3.6\,kpc
distance). On the other hand, if the characteristic age of 117~kyr is
correct, we probably would not expect to see a supernova remnant,
while assuming a younger age we would not expect a predominantly
thermal X-ray emission, with no magnetospheric component as seen for
younger pulsars.

A third possibility might be that the nebula around \rrat\, gets
additional power from the large magnetic energy of this object through
mechanisms such as ambipolar diffusion which, instead of going into
powering the X-ray emission of this object (as in the magnetar case;
Thompson \& Duncan 1995), releases its energy in the pulsar wind, or
through repeated and powerful transient outbursts (Ibrahim et
al. 2004; Halpern et al. 2008; Muno et al. 2007). So far, no
magnetically-powered nebula has been detected around magnetars. Many
observational biases might have prevented their detection, such as
magnetars' large distances, and their very bright emission which
usually cannot be observed with high-spatial resolution, but only
using 1-dimensional modes (aka timing modes). However, in a very few
cases there has been hints for extended emission around magnetars (see
e.g. Patel et al. 2003). Beside \rrat\, there are seven
rotation-powered pulsars with magnetic fields greater than the quantum
critical field, and two of them, PSR\,J1846--0258 and
PSR\,J1119--6127, have PWNe associated (Helfand, Collins \& Gotthelf
2003; Gonzalez \& Safi-Harb~2003). They have also very high spin-down
power ($\dot{E}_{\rm rot} > 10^{36}$\ergs ), making difficult any
speculation on the possible magnetic contribution to their
nebulae. However, note that PSR\,J1846--0258 did show episodes of
magnetar-like activity (Kumar \& Safi-Harb 2008; Gavriil et al. 2008),
and strong variability has been observed in its PWN (Ng et al. 2009)
in coincidence with its outburst, which supports the hypothesis of
magnetic energy as an additional energy source to power its PWN. For
the remaining five highly magnetic pulsars, it is possible that
magnetic-powered nebulae are commonplace, and that selection effects
(e.g. the large distances) have precluded us detecting them.

In summary, given the low number of counts we cannot give a conclusive
answer on the nature of this extended X-ray emission observed around
\rrat. It is clear, however, that with our current picture of PWNe
around pulsars, and with the information we now have on \rrat, all
traditional interpretations fail at explaining this extended
emission. Currently the most viable interpretation seems to be the
presence of magnetic contribution to powering of this unusual
nebula. Further monitoring observations of \rrat\, are crucial to
constrain any variability of this extended emission, which could be
due to episodic magnetar-like outbursts, as in the case of
PSR\,J1846--0258.

\acknowledgements

We wish to thank T.~Aldcroft and P.~Plucinsky for checking the \CXO\,
aspect reconstruction during this observation and for useful
suggestions, and K.~Borkowski, O.~Kargaltsev and P.~Esposito for
comments.  NR acknowledges support from an NWO Veni Fellowship. MAM is
supported by a WV EPSCoR grant, and a SAO guest investigator grant.

\begin{center}
\begin{figure*}[t]
\hbox{
\includegraphics[height=6.5cm]{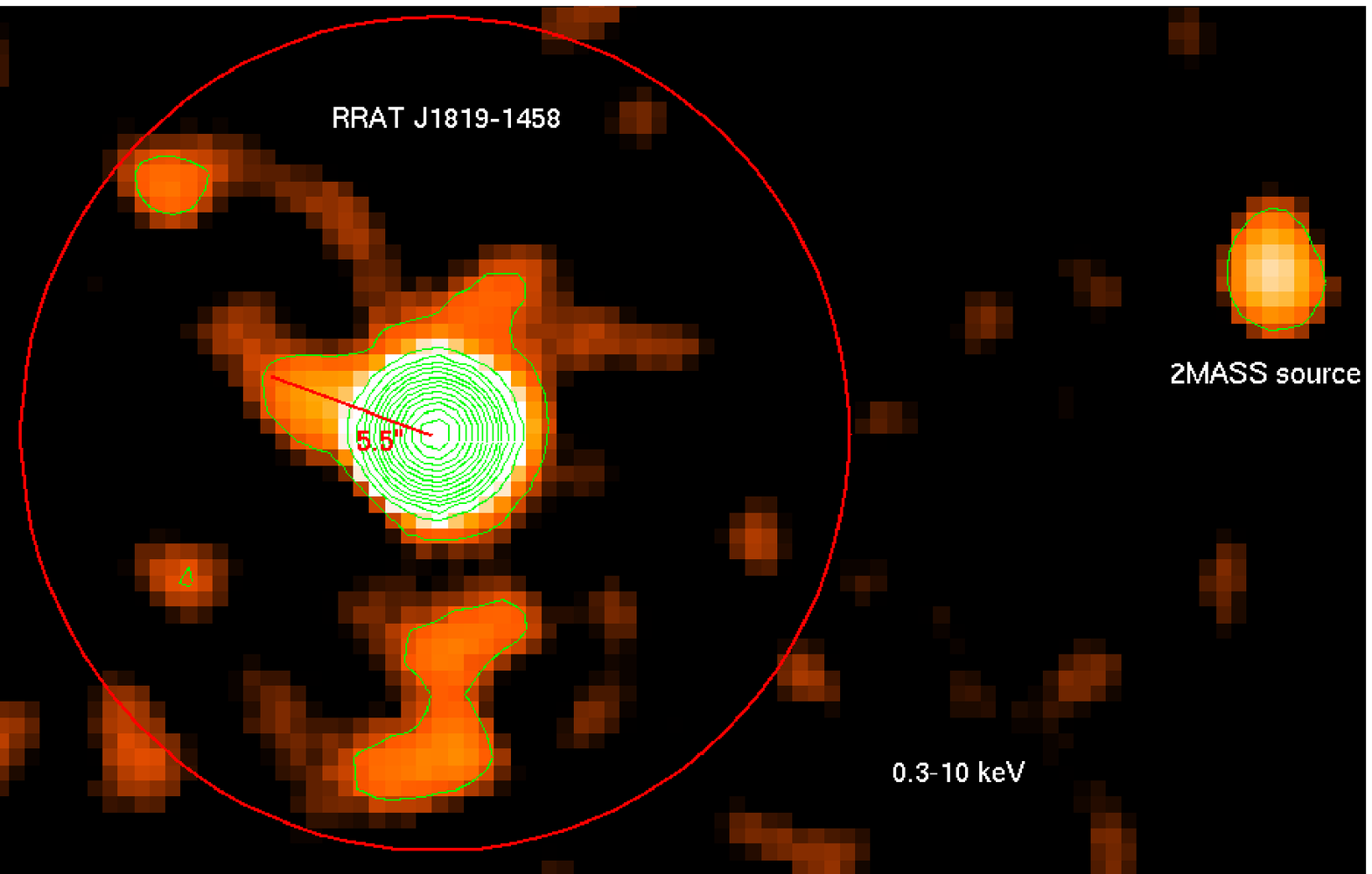}
\hspace{0.8cm}
\includegraphics[height=6.5cm]{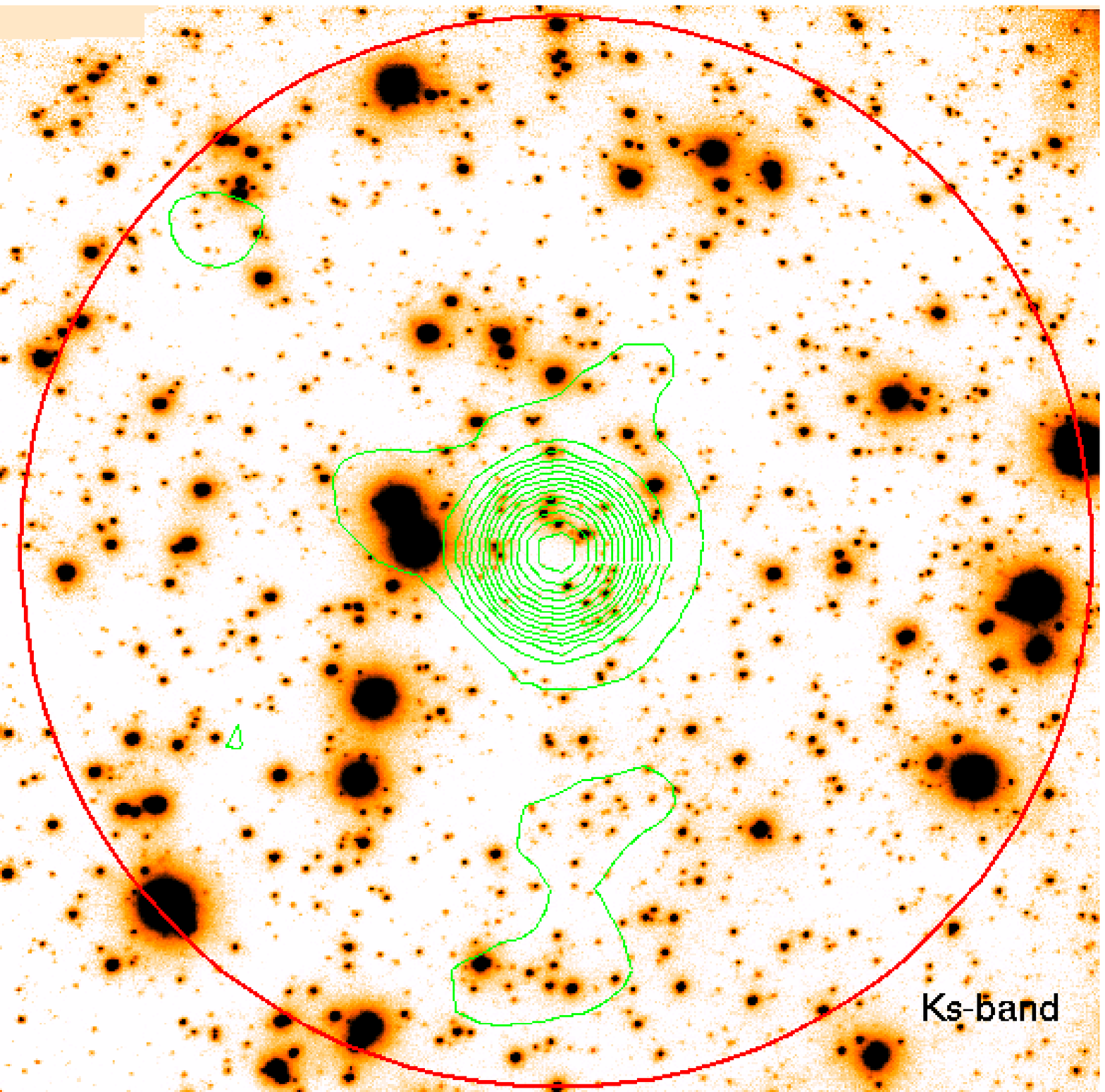}}
\caption{{\em Left Panel}: 0.3--10\,keV log-image of our 30\,ks \CXO\,
  ACIS-S observation of \rrat, with a circular region of 13\arcsec\,
  over-plotted, and the contours of the extended emission (from
  3$\sigma$ increasing by 1$\sigma$ each). The image has been smoothed
  with a Gaussian function with a radius of 3 pixels. {\em Right
    panel}: VLT--NACO image in the K$_s$-band of the field of \rrat\,
  (Rea et al. 2009) with over-plotted the same circular region and
  contours as in the left panel. North is top, and East is left.}
\end{figure*}
\end{center}

\begin{center}
\begin{figure*}
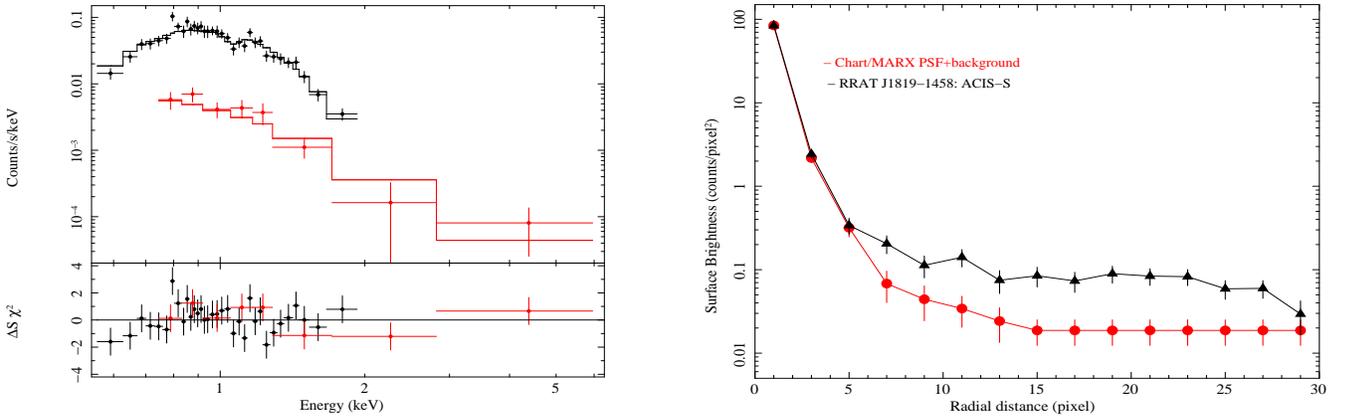

\hbox{
\includegraphics[height=8cm, width=5.5cm,angle=-90]{f2a.eps}
\hspace{1cm}
\includegraphics[height=8.5cm, width=5.5cm,angle=-90]{f2b.eps}}
\caption{{\em Left panel}: In black we show the ACIS-S spectrum of
  \rrat\, modeled with an absorbed blackbody plus a 1\,keV absorption
  line, while in red we show the spectrum of the extended emission
  fitted with a power-law. {\em Right panel}: surface brightness of
  the background-subtracted ACIS-S image of \rrat\, (black) and of the
  {\it Chart/MARX } PSF plus a constant background (red). One ACIS-S
  pixel corresponds to 0.492\arcsec.}
\end{figure*}
\end{center}

\begin{figure}[t]
\centerline{\includegraphics[width=7cm,height=7.5cm,angle=-90]{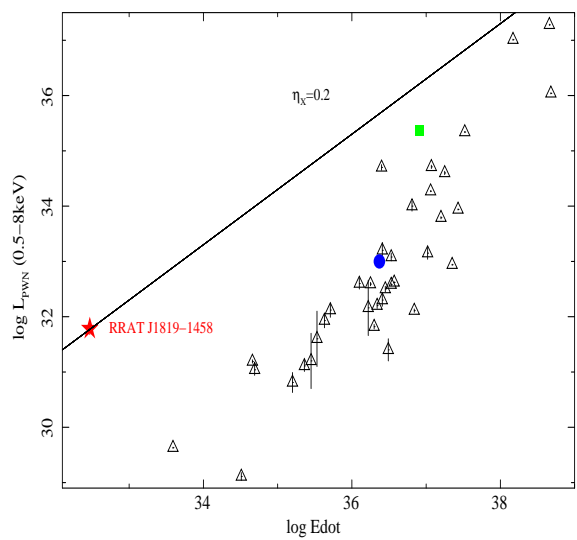}}
\caption{Luminosity of all the known X-ray PWNe (in the 0.5--8\,keV
  energy range) compared to the rotational power of the hosted
  pulsars. All the data, except for \rrat\,  (this work) and PSR\,J1846-0258 (revised distance, Leahy \& Tian 2008), were taken from Kargaltsev \& Pavlov (2008). The green square
  and blue circle report on the high-B pulsars PSR\,J1846--0258 and
  PSR\,J1119--6127, respectively. The solid line represents
  $\eta_{X}=0.2$. Errors in the pulsar distances are not taken into account in
  this plot.}
\end{figure}

\end{document}